\documentclass{article}

\usepackage{arxiv}
\usepackage{float}
\usepackage[utf8]{inputenc} 
\usepackage[T1]{fontenc}    
\usepackage{hyperref}       
\usepackage{url}            
\usepackage{booktabs}       
\usepackage{amsfonts}       
\usepackage{nicefrac}       
\usepackage{microtype}      
\usepackage{lipsum}
\usepackage{graphicx}
\graphicspath{ {./images/} }
\usepackage{titlesec}
\usepackage{caption}
\usepackage{array}
\usepackage{geometry}
\usepackage[spanish]{babel}

\addto\captionsspanish{}

\title{Modelado y Gemelos Digitales en el contexto fotovoltaico}

\author{
 Franco Bertani Matung [0009-0003-0039-7239] \\
  Facultad de Ingeniería\\
  Universidad Nacional de Cuyo\\
  Modenza POBOX 5502 - Argentina \\
  \texttt{bertani.franco@uncuyo.edu.ar} \\
   \And
 Juan Cruz Esquembre Santamaría [0009-0002-8219-0524]\\
  Facultad de Ingeniería\\
  Universidad Nacional de Cuyo\\
  Modenza POBOX 5502 - Argentina \\
  \texttt{esquembre.juan@uncuyo.edu.ar} \\
  \And
 Ricardo R. Palma [0000-0002-1864-7625] \\
  Instituto de Ingeniería Industrial \\
  Universidad Nacional de Cuyo\\
  Ciudad Universitaria Mendoza POBOX 5502 Argentina \\
  \texttt{rpalma@uncu.edu.ar} \\
   \And
 Fabricio Orlando Sanchez Varretti [0000-0002-7239-9562]  \\
  Universidad Tecnológica Nacional \\
  Facultad Regional San Rafael\\
  CONICET, SiCo, Mendoza, Argentina \\
  \texttt{fsanchez@frsr.utn.edu.ar} \\ 
}

\begin{document}
\maketitle
\vspace{3cm}
\begin{abstract}
La industria fotovoltaica enfrenta el desafío de optimizar el rendimiento y la gestión de sus sistemas en un entorno cada vez más digitalizado. En este contexto, los digital twins o gemelos digitales ofrecen una solución innovadora: modelos virtuales que replican en tiempo real el comportamiento de instalaciones solares. Esta tecnología permite anticipar fallos, mejorar la eficiencia operativa y facilitar la toma de decisiones basada en datos. El presente informe analiza su aplicación en el sector fotovoltaico, destacando sus beneficios y potencial transformador.
\end{abstract}

\section{Introducción}
\vspace{1cm}
En los últimos años se ha visto en la Argentina una necesidad creciente de renovar la matriz energética. Impulsado principalmente por una demanda de energía en constante crecimiento y la toma de conciencia de técnicas y procesos más sustentables en la industria.  Esto ha dado lugar a un interés en constante aumento de las energías renovables como solución a esta “renovación”, para garantizar la sostenibilidad y mitigar los efectos del cambio climático.\cite{ministerio_de_relaciones_exteriores_comercio_internacional_y_culto_produccion_2024}

Entre las distintas  fuentes renovables, la energía solar ha ganado importancia debido a su abundancia y disponibilidad en prácticamente todas las regiones del mundo. El aprovechamiento de la energía solar,principalmente en sistemas fotovoltaicos,se ha expandido a una escala global, convirtiéndose en una herramienta fundamental en la renovación de la matriz energética: la capacidad instalada global superó el teravatio en 2022, y se proyecta que alcance los 2,3 TW para 2025.\cite{solarpower_europe_global_2022} Sin embargo, a pesar de su madurez tecnológica y reducción sostenida de costos, el diseño y operación óptima de sistemas solares fotovoltaicos presentan desafíos. Uno de ellos es la incertidumbre respecto a la estimación de su rendimiento real.\cite{thevenard_estimating_2013}

La variabilidad climática, el impacto de las sombras, las pérdidas por uso y las ineficiencias de los componentes son algunos de los factores que dificultan predecir con exactitud la generación eléctrica. Esta falta de certeza puede verse reflejada en sobredimensionamiento, baja eficiencia, o una subestimación del retorno económico del proyecto. En consecuencia, contar con herramientas de simulación robustas y precisas es clave para asegurar el éxito técnico y financiero de las instalaciones fotovoltaicas.
\cite{paletta_eclipse_2022}
La sustentabilidad debe ser más que una meta aspiracional y debe basarse en hechos concretos y resultados medibles. Es necesario contar con herramientas que no solo optimicen el diseño de sistemas fotovoltaicos, sino que también proporcionen resultados fiables y precisos.

En este escenario,podemos ver una evolución en las herramientas de modelado energético: los llamados gemelos digitales (Digital Twins). A diferencia de los modelos tradicionales, que representan el sistema de forma estática (Digital Model), o que simplemente reciben datos reales sin retroalimentar el sistema (Digital Shadow), un gemelo digital establece una interacción bidireccional con el sistema físico, permitiendo ajustar el comportamiento del modelo en tiempo real a partir de datos operacionales y técnicas de inteligencia artificial.\cite{luis_mollineda_digital_2024}

Esta nueva generación de herramientas de industria 4.0  no solo nos permite simular distintos escenarios sino también prevenir fallos, optimizar el mantenimiento y predecir la producción energética con mayor exactitud\cite{afridi_artificial_2021}.
Este trabajo tiene como objetivo hacer un relevamiento de los distintos tipos de modelos en el contexto fotovoltaico y analizar la madurez digital de distintos simuladores fotovoltaicos para evaluar su potencial de integración en una arquitectura de gemelo digital, tomándose como referencia un caso de estudio real.
\renewcommand{\baselinestretch}{1.5}
\section{Marco Teórico}
\subsection{Historia del concepto de gemelo digital}
El término gemelo digital (digital twin) fue introducido por Michael Grieves en 2002 en el contexto de la gestión del ciclo de vida del producto (Product Lifecycle Management, PLM), y desde entonces ha evolucionado hasta convertirse en una de las tecnologías emergentes más prometedoras para el monitoreo, control y optimización de sistemas físicos complejos.\cite{grieves_intelligent_2022}

Un gemelo digital puede definirse como una representación virtual dinámica de un sistema físico, conectada en tiempo real mediante flujos de datos que permiten analizar su estado, predecir su comportamiento futuro y retroalimentar su operación. En el caso de instalaciones energéticas, su implementación permite reducir costos operativos, mejorar la eficiencia y anticipar fallos, lo que resulta especialmente útil en sistemas expuestos a condiciones climáticas cambiantes, como los fotovoltaicos.\cite{angelova_review_2024}

Diversos autores han propuesto clasificaciones del grado de madurez de estas representaciones digitales, existen tres niveles principales como los que se visualizan a continuación\cite{hamid_enhancing_2025}:

•	Digital Model (DM): Modelo desconectado del sistema físico. Representa el diseño o la configuración esperada del sistema, sin interacción con datos reales.

•	Digital Shadow (DS): Modelo que recibe datos reales en tiempo casi real, pero sin enviar retroalimentación al sistema físico.

•	Digital Twin (DT): Modelo completamente conectado en ambas direcciones. Puede recibir datos del entorno y del sistema, procesarlos, realizar predicciones y enviar acciones o recomendaciones al sistema físico.
\begin{figure}[H]
    \centering
    \includegraphics[width=0.75\linewidth]{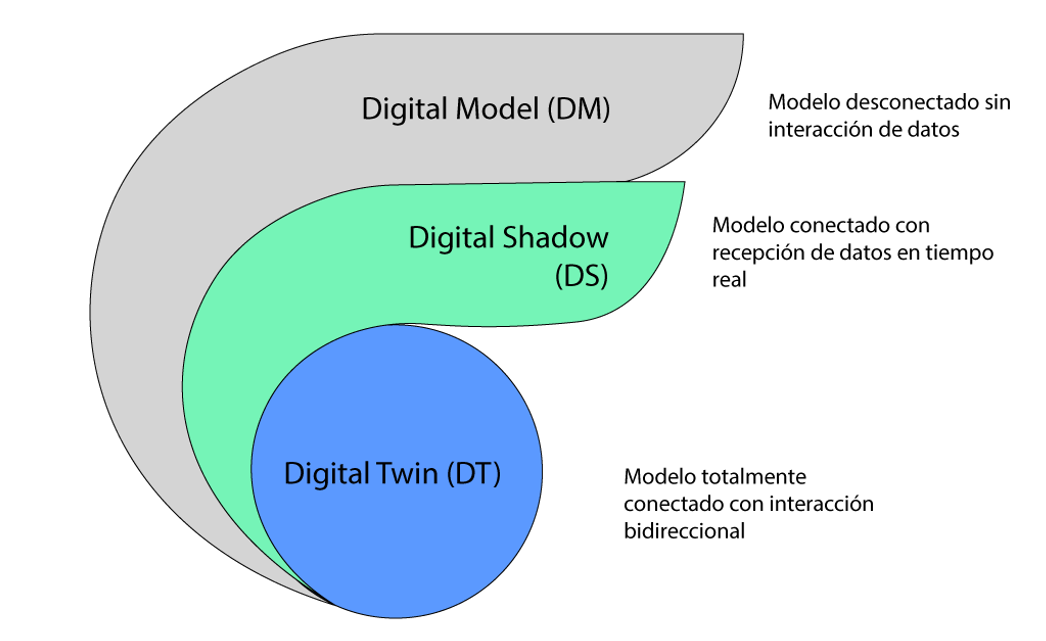}
    \caption{Niveles de simulación digital}
    \captionsetup[figure]{name={Fig.},}
    \label{fig:enter-label}
\end{figure}

En lo que sigue se representa una matriz categórica donde se posicionan distintos tipos de modelos digitales según su grado de integración de datos (eje horizontal) y su nivel de conectividad (eje vertical), permitiendo visualizar la evolución hacia un gemelo digital completo.

\begin{figure}[H]
    \centering
    \includegraphics[width=0.75\linewidth]{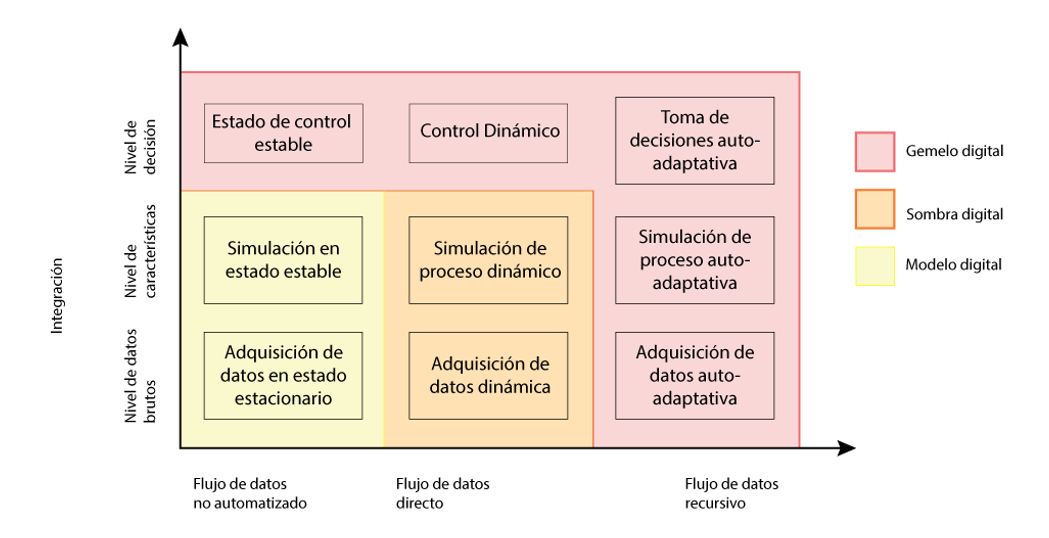}
    \caption{Integración vs Conectividad en el simulado digital}
    \label{fig:enter-label}
\end{figure}
\renewcommand{\baselinestretch}{1.5}
\subsection{Aplicación de gemelos digitales en sistemas fotovoltaicos}
La integración de gemelos digitales en el sector energético, y en particular en sistemas solares fotovoltaicos, representa una oportunidad para mejorar la precisión del modelado, la gestión operativa y la eficiencia de producción.Los gemelos digitales aplicados a pequeñas y medianas instalaciones fotovoltaicas permiten realizar predicciones de generación (nowcasting) basadas exclusivamente en datos medidos en sitio, sin necesidad de información técnica completa del sistema.\cite{guzman_razo_genetic_2023}

Las ventajas clave de la aplicación de gemelos digitales en sistemas fotovoltaicos incluyen la posibilidad de realizar un monitoreo en tiempo real del estado operativo y energético del sistema, lo que permite detectar anomalías y optimizar el funcionamiento de manera continua. Además, habilitan la simulación dinámica de escenarios de falla, evaluando el impacto de distintas condiciones operativas antes de que ocurran. Otra característica es la capacidad de auto-ajustar el modelo digital en función de las condiciones climáticas reales, como pueden ser variaciones en la irradiancia o la temperatura ambiente. Finalmente, estas herramientas permiten implementar estrategias de mantenimiento predictivo, al identificar desviaciones entre el comportamiento esperado (modelo) y el comportamiento real del sistema, reduciendo tiempos de inactividad y aumentando la eficiencia global\cite{hamid_enhancing_2025}.
Estudios recientes también han demostrado que frameworks como ROS2 o arquitecturas abiertas basadas en Python pueden usarse para construir gemelos digitales solares escalables, utilizando nodos para variables meteorológicas y energéticas.\cite{diaz_rivera_gemelo_2023}.
\subsection{Simuladores fotovoltaicos y su clasificación digital}
En el ámbito fotovoltaico, existen múltiples simuladores que permiten estimar la producción energética de un sistema en función de condiciones climáticas, técnicas y configuraciones específicas. Sin embargo, no todos estos simuladores pueden considerarse "gemelos digitales". A continuación se describe su clasificación según la categorización DM–DS–DT \cite{urupira_tafadzwa__rix_arnold_pdf_2017}:
\begin{table}[ht]
  \caption{Comparación entre IRELIA, PVsyst y SAM según el nivel de madurez digital. Se detalla el tipo de modelo (DM, DS, DT), sus características principales y las limitaciones para alcanzar una simulación basada en gemelos digitales.\cite{kumar_design_2021}}
  \centering
  \renewcommand{\arraystretch}{1.2}
  \begin{tabular}{>{\raggedright\arraybackslash}p{2.5cm} 
                  >{\raggedright\arraybackslash}p{3cm} 
                  >{\raggedright\arraybackslash}p{5.5cm} 
                  >{\raggedright\arraybackslash}p{5.5cm}}
    \toprule
    \textbf{Simulador} & \textbf{Nivel Digital} & \textbf{Características clave} & \textbf{Limitaciones} \\
    \midrule
    IRELIA & DM (Modelo Digital) & Basado en scripts, orientado a docencia e investigación. Modela configuraciones sin conexión con datos reales. & No recibe datos en tiempo real. No realiza retroalimentación. \\
    PVsyst & DS (Sombra Digital) & Simulador técnico de alta precisión. Permite importar bases meteorológicas y modelar sombreados, pérdidas e inclinaciones. & No se actualiza dinámicamente con datos operativos. \\
    SAM & Entre DS y DT & Simulación técnica y económica. Integra bases reales (NSRDB), scripting en Python/Matlab y exportación de resultados. & Su transformación en DT depende de sensores y plataformas externas. \\
    \bottomrule
  \end{tabular}
  \label{tab:comparacion_simuladores}
\end{table}

El simulador IRELIA se clasifica dentro del nivel digital DM (Modelo Digital). Está diseñado principalmente para fines educativos y de investigación, y se basa en el uso de scripts para modelar configuraciones fotovoltaicas. Sin embargo, no cuenta con conexión a datos reales ni capacidades de retroalimentación, lo que limita su aplicación en contextos operativos dinámicos.

Por otro lado, PVsyst se ubica en el nivel DS (Sombra Digital). Es un simulador técnico de alta precisión que permite importar bases meteorológicas y modelar aspectos como sombreados, pérdidas e inclinaciones. A pesar de su robustez técnica, no se actualiza dinámicamente con datos operativos, lo que impide su uso como gemelo digital en tiempo real.

Finalmente, el simulador SAM (System Advisor Model) se encuentra en una posición intermedia entre los niveles DS y DT (Gemelo Digital). Ofrece capacidades tanto técnicas como económicas, y permite la integración con bases de datos reales como NSRDB. Además, admite scripting en Python o Matlab y la exportación de resultados para análisis posterior. No obstante, para alcanzar plenamente el nivel de gemelo digital, requiere ser complementado con sensores y plataformas externas que habiliten la adquisición de datos en tiempo real y la retroalimentación del sistema.

Como se puede observar, ningún simulador en su configuración base alcanza plenamente el nivel de gemelo digital. No obstante, mediante la integración con plataformas de monitoreo, APIs de adquisición de datos, y algoritmos de inteligencia artificial, es posible extender su funcionalidad y migrar hacia esquemas más inteligentes y adaptativos \cite{angelova_review_2024}

\subsection{Modelos híbridos y Machine learning.}
El paso de un modelo digital estático a un gemelo digital funcional suele requerir la incorporación de algoritmos de machine learning para mejorar las capacidades de predicción y adaptación. Como se detalla en la literatura, técnicas como redes neuronales artificiales, random forest y modelos autorregresivos han demostrado buenos resultados al predecir producción energética en condiciones dinámicas\cite{uddin_machine_2024} . Un enfoque híbrido, que combine modelos físicos como los ofrecidos por SAM y PVsyst con componentes de aprendizaje automático, permite capturar tanto el comportamiento teórico como las desviaciones observadas en la operación real, lo cual es fundamental para el desarrollo de gemelos digitales eficientes.

\section{Metodología}
\subsection{Enfoque general}
El enfoque general de este trabajo se basa en una metodología de análisis comparativo y exploratorio, cuyo objetivo principal es evaluar el grado de madurez digital de dos herramientas de simulación fotovoltaica: PVsyst y SAM. Esta metodología se estructura en cuatro etapas. En primer lugar, se realiza una caracterización funcional, donde se definen las capacidades específicas de cada simulador. Luego, se lleva a cabo la simulación de una planta fotovoltaica, utilizando ambos softwares para estimar su rendimiento bajo condiciones determinadas. Posteriormente, se procede a la comparación de resultados, contrastando los datos obtenidos en las simulaciones con datos reales de operación. Finalmente, se realiza una evaluación de integración, en la que se analiza el potencial de cada herramienta para incorporarse dentro de una arquitectura de gemelo digital.
La metodología incluye:
\begin{figure}[H]
    \centering
    \includegraphics[width=1\linewidth]{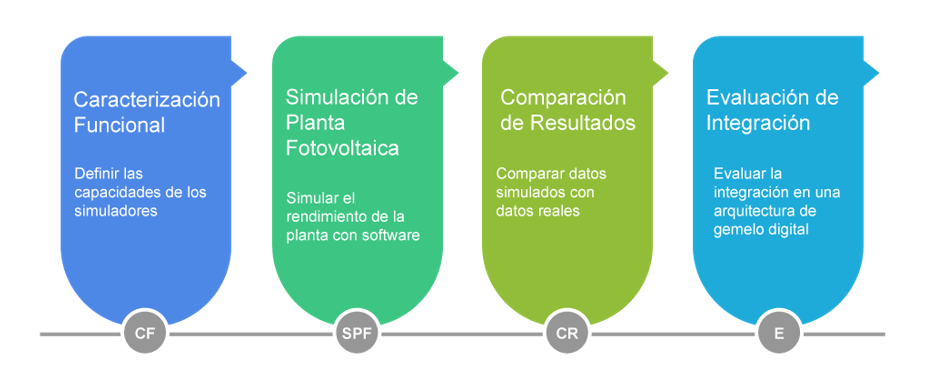}
    \caption{Metodología de trabajo}
    \label{fig:enter-label}
\end{figure}
\subsection{Caso de estudio: planta fotovoltaica de referencia}
El análisis se basa en una instalación fotovoltaica ubicada en el distrito de Elmah, provincia de Antalya, Turquía\cite{sancar_comparative_2023}. Se trata de un sistema conectado a red, instalado sobre el techo de un almacén frigorífico y consta de 360 paneles fotovoltaicos monocristalinos distribuidos en dos orientaciones con una inclinación de 10°. Cada orientación está asociada a un inversor, sumando un total de dos inversores de 50 kW cada uno. Los paneles tienen una potencia máxima de 325 Wp, un voltaje Vmpp de 33.68 V, y una corriente máxima Impp de 9.65 A. El sistema opera conectado a la red, con un diseño optimizado para reducir pérdidas por sombras y maximizar la captación energética según las condiciones locales.

\begin{figure}[H]
    \centering
    \includegraphics[width=0.7\linewidth]{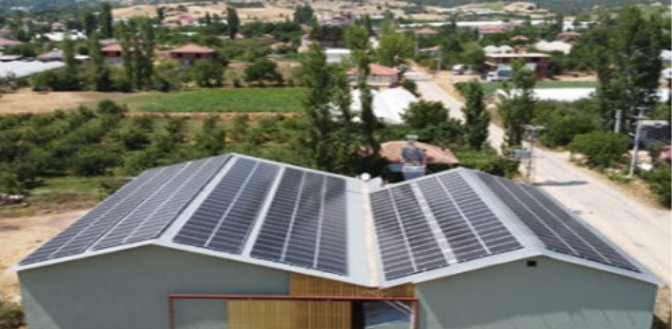}
    \caption{Instalación fotovoltaica ubicada en el distrito de Elmah}
    \label{fig:enter-label}
\end{figure}
La caracterización técnica del sistema se complementa con una tabla de propiedades de los paneles, donde se detallan aspectos clave como el número total de módulos (360), su distribución en 18 cadenas de 20 paneles cada una, y parámetros eléctricos como la potencia máxima de 325 Wp, el voltaje en el punto de máxima potencia (Vmpp) de 33.68 V y la corriente correspondiente (Impp) de 9.65 A. Estos valores permiten dimensionar con precisión el comportamiento eléctrico del sistema y sirven como base para las simulaciones y análisis comparativos posteriores.
\begin{table}[ht]
  \caption{Propiedades técnicas del sistema fotovoltaico. Se detallan los parámetros eléctricos de los paneles, así como la configuración general del sistema.}
  \centering
  \renewcommand{\arraystretch}{1.2}
  \begin{tabular}{>{\raggedright\arraybackslash}p{8cm} >{\raggedright\arraybackslash}p{4cm}}
    \toprule
    \textbf{Propiedad} & \textbf{Valor} \\
    \midrule
    Número de Cadenas en el Sistema & 18 \\
    Número de Paneles por Cadena & 20 \\
    Número Total de Paneles & 360 \\
    Tipo de Panel & Panel Monocristalino \\
    Ángulo del Panel Este-Oeste & 10º \\
    Potencia Máxima & 325 Wp \\
    Voltaje Máximo (Vmpp) & 33.68 V \\
    Corriente Máxima (Impp) & 9.65 A \\
    Voltaje de Circuito Abierto (Voc) & 40.55 V \\
    Corriente de Cortocircuito (Isc) & 10.26 A \\
    \bottomrule
  \end{tabular}
  \label{tab:propiedades_paneles}
\end{table}

En cuanto a las propiedades mecánicas de los paneles, se trata de módulos monocristalinos compuestos por 60 células, con dimensiones de 1670 mm por 1000 mm y un espesor de 35 mm. Cada panel tiene un peso aproximado de 18 kg, con una tolerancia de ±1 kg. El vidrio frontal es templado de 3.2 mm, con alta transmitancia y bajo contenido de hierro, mientras que la parte trasera está protegida por una película de poliéster (PET). El marco es de aluminio anodizado, y la caja de conexiones incorpora tres diodos de derivación, con conectores compatibles con el estándar MC4.

\begin{table}[ht]
  \caption{Propiedades mecánicas de los paneles solares, detallando dimensiones, materiales y componentes principales.}
  \centering
  \renewcommand{\arraystretch}{1.2}
  \begin{tabular}{>{\raggedright\arraybackslash}p{7.5cm} >{\raggedright\arraybackslash}p{5.5cm}}
    \toprule
    \textbf{Propiedad} & \textbf{Valor} \\
    \midrule
    Dimensiones del Panel & 1670mm x 1000mm x 35mm \\
    Célula & Monocristalina \\
    Número de Células & 60 \\
    Peso & 18kg ± 1kg \\
    Vidrio Frontal & Vidrio templado de 3.2mm con alta transmitancia y bajo contenido de hierro \\
    Protección Trasera & Película de Poliéster PET \\
    Marco & Aluminio Anodizado \\
    Caja de Conexiones & 3 Diodos de Derivación \\
    Conector & Compatible con MC4 \\
    \bottomrule
  \end{tabular}
  \label{tab:propiedades_mecanicas_paneles}
\end{table}

En lo que respecta a los inversores utilizados en la planta, se emplean dos unidades con una potencia nominal de 50 kW cada una. Cada inversor gestiona 9 series de paneles y cuenta con tres seguidores MPPT. El voltaje máximo de entrada en corriente continua (DC) es de 1000 V, con un voltaje de arranque de 420 V y un rango operativo MPPT entre 480 y 800 V. La corriente máxima de cortocircuito que pueden manejar es de 165 A. En cuanto a la salida en corriente alterna (AC), la potencia nominal es de 50 kW, con una potencia máxima de 55 kW, un voltaje de salida de 400/230 V y una corriente máxima de 80 A.

\begin{table}[ht]
  \caption{Propiedades técnicas del inversor utilizado en la planta de Elmali, incluyendo parámetros eléctricos y configuraciones relevantes.}
  \centering
  \renewcommand{\arraystretch}{1.2}
  \begin{tabular}{>{\raggedright\arraybackslash}p{8cm} >{\raggedright\arraybackslash}p{4cm}}
    \toprule
    \textbf{Propiedad} & \textbf{Valor} \\
    \midrule
    Número de Inversores Utilizados & 2 \\
    Número de Series por Inversor & 9 \\
    Potencia Nominal del Inversor & 50 kW \\
    Voltaje de Entrada Máximo DC & 1000 V \\
    Voltaje de Arranque DC & 420 V \\
    Rango de Voltaje de Arranque MPTT DC & 480--800 V \\
    Número MPTT & 3 \\
    Corriente de Entrada Máxima & 108 A \\
    Corriente Máxima de Cortocircuito & 165 A \\
    Potencia de Salida Nominal AC & 50 kW \\
    Potencia de Salida Máxima AC & 55 kW \\
    Voltaje AC & 400 / 230 V \\
    Flujo Máximo de Corriente AC & 80 A \\
    Frecuencia & 50 Hz / 60 Hz \\
    \bottomrule
  \end{tabular}
  \label{tab:propiedades_inversor_elmali}
\end{table}

Para las simulaciones, se utilizaron datos meteorológicos del recurso NSRDB en SAM y se importaron al entorno de PVsyst para garantizar la consistencia en la comparación. Las configuraciones específicas, como ángulos de inclinación y orientaciones de los paneles, fueron replicadas en ambos programas para evaluar diferencias en estimaciones de producción y pérdidas energéticas.

\subsection{Procedimiento de simulación}

Se configuraron ambos sistemas en los simuladores PVsyst y SAM bajo condiciones de diseño equivalentes, replicando parámetros como la geometría del sistema, la inclinación y orientación de los paneles, así como las pérdidas estimadas. Para asegurar una base común de comparación, se empleó el mismo conjunto de datos meteorológicos (NSRDB) y se mantuvo la coherencia en variables clave como el número de cadenas, el tipo de panel y la eficiencia de los inversores.

A modo complementario, se incluyen a continuación capturas de pantalla que documentan las configuraciones realizadas en cada simulador. Estas imágenes permiten visualizar directamente los valores ingresados y las características técnicas consideradas, lo que refuerza la transparencia del proceso y la validez del análisis comparativo.

La imagen presentada corresponde a la sección de ingreso de variables climáticas en PVsyst. En esta interfaz se introducen datos mensuales como la irradiación global y difusa en plano horizontal, la temperatura media, la velocidad del viento, la turbidez atmosférica (Linke) y la humedad relativa. Estos parámetros son fundamentales para caracterizar con precisión el recurso solar disponible en el sitio de estudio y asegurar que las simulaciones reflejen adecuadamente las condiciones ambientales locales a lo largo del año.
\begin{figure}[H]
    \centering
    \includegraphics[width=1\linewidth]{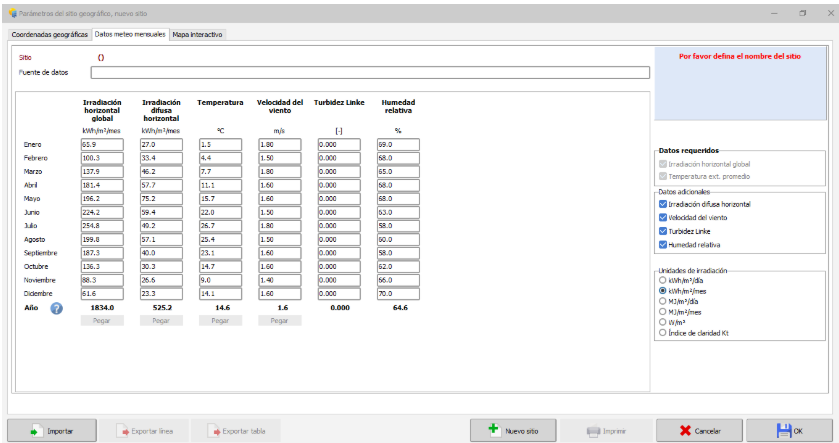}
    \caption{Proceso de data entry de variables climáticas en Pv-syst.}
    \label{fig:enter-label}
\end{figure}

La imagen siguiente corresponde a la configuración del campo fotovoltaico en PVsyst, donde se especifican las orientaciones múltiples del sistema. En este caso, se definieron dos orientaciones con una inclinación de 10° y ángulos azimutales de 53° y -127°, respectivamente. Esta configuración permite representar con mayor precisión la disposición real de los paneles en el sitio. Además, se incluyen parámetros de optimización relacionados con la irradiación anual y estacional, así como indicadores clave como el factor de transposición (FT) y la irradiación global en el plano del colector, lo que contribuye a evaluar el rendimiento energético del sistema bajo distintas condiciones de orientación.

\begin{figure}[H]
    \centering
    \includegraphics[width=1\linewidth]{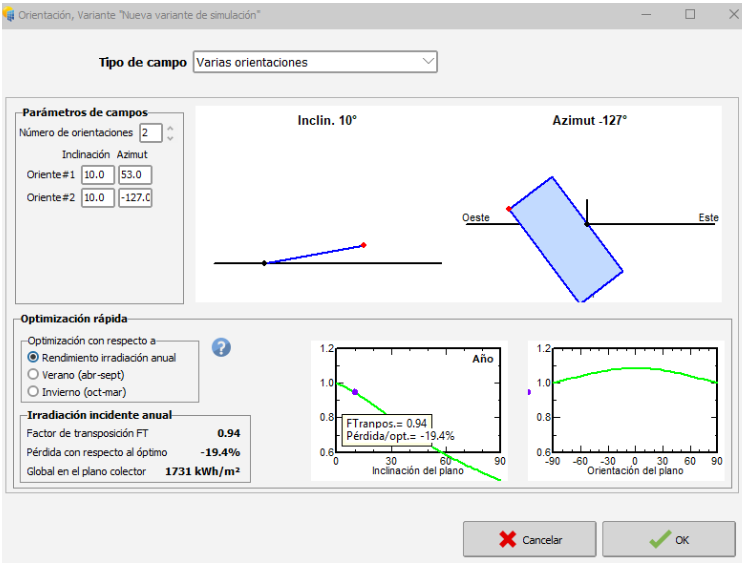}
    \caption{Data entry de las características físicas en Pv-syst.}
    \label{fig:enter-label}
\end{figure}

La siguiente captura corresponde a la configuración detallada del sistema fotovoltaico en PVsyst, donde se definen aspectos clave como la selección del módulo fotovoltaico y del inversor, así como la orientación e inclinación de los subconjuntos. Se observa la elección del modelo de panel AEG-AS-M1208-HZ-S25 y del inversor CSI-S6XT2-GFL, junto con parámetros técnicos como el voltaje en circuito abierto (Voc), la potencia planeada y la proporción de potencia DC respecto a la nominal. Esta sección permite validar que la configuración del sistema refleja fielmente las condiciones reales del proyecto.

\begin{figure}
    \centering
    \includegraphics[width=1\linewidth]{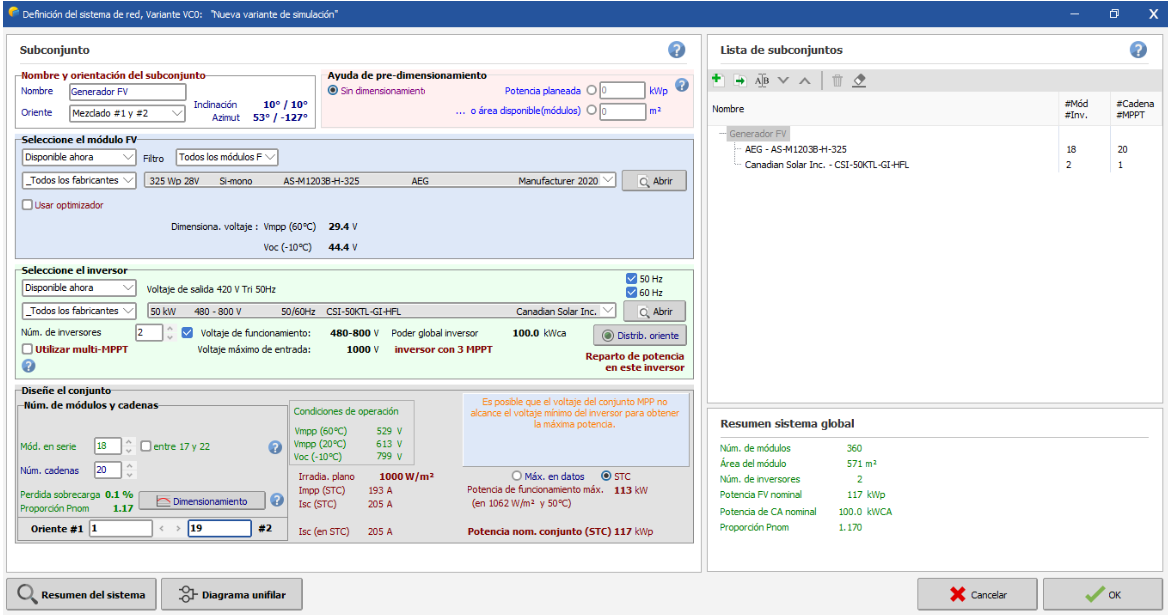}
    \caption{Carga de las características técnicas de los paneles }
    \label{fig:enter-label}
\end{figure}

La siguiente imágen corresponde a la interfaz principal del entorno de simulación en PVsyst, donde se gestionan los parámetros generales del proyecto. En esta sección se definen elementos como el nombre del proyecto, el archivo del sitio, los datos meteorológicos y la variante de simulación utilizada. También se accede a las opciones de ejecución de la simulación y visualización de resultados avanzados. A la derecha, se presenta un resumen con indicadores clave de rendimiento del sistema, como la producción específica, las pérdidas y la proporción de rendimiento, lo que permite una evaluación preliminar del comportamiento energético del sistema modelado.

\begin{figure}[H]
    \centering
    \includegraphics[width=1\linewidth]{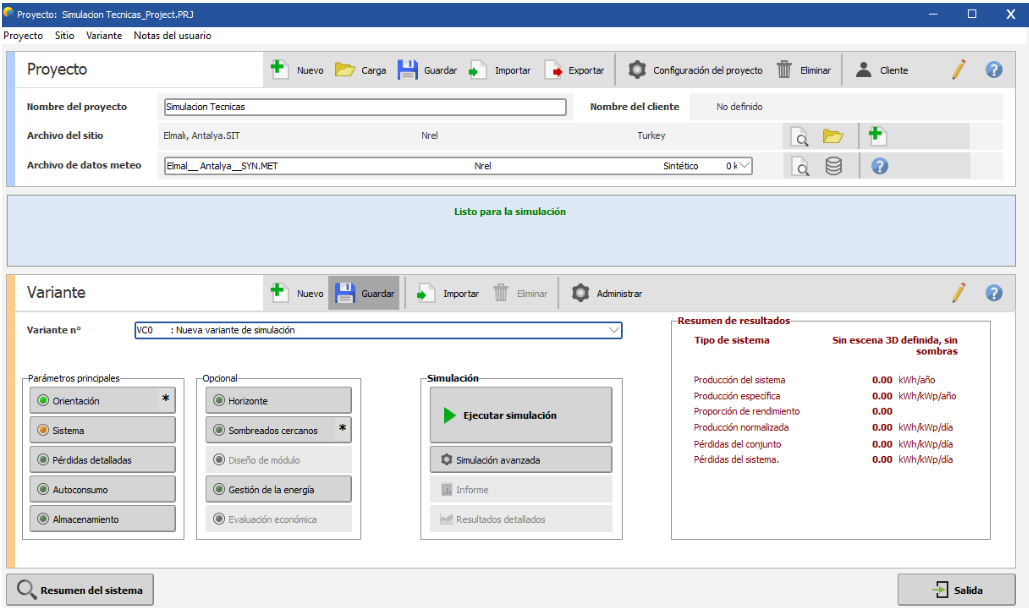}
    \caption{Interfaz de Pv-syst}
    \label{fig:enter-label}
\end{figure}

La siguiente imagen corresponde a la interfaz del software SAM, específicamente en la sección de selección de módulos fotovoltaicos mediante el modelo de rendimiento CEC. En ella se visualiza la base de datos de módulos disponibles, junto con sus principales características técnicas como tecnología, eficiencia, dimensiones y potencia nominal. Se destaca la selección del módulo WAAREE WSM-325, cuyas especificaciones detalladas se muestran a la derecha, incluyendo parámetros como la eficiencia nominal (19.7 porcentual), la potencia máxima (325W), el voltaje en el punto de máxima potencia (Vmp) y la corriente correspondiente (Imp). Además, se incluye una gráfica de comportamiento eléctrico bajo condiciones estándar, lo que permite validar la coherencia del modelo seleccionado con respecto al sistema real.

\begin{figure}[H]
    \centering
    \includegraphics[width=1\linewidth]{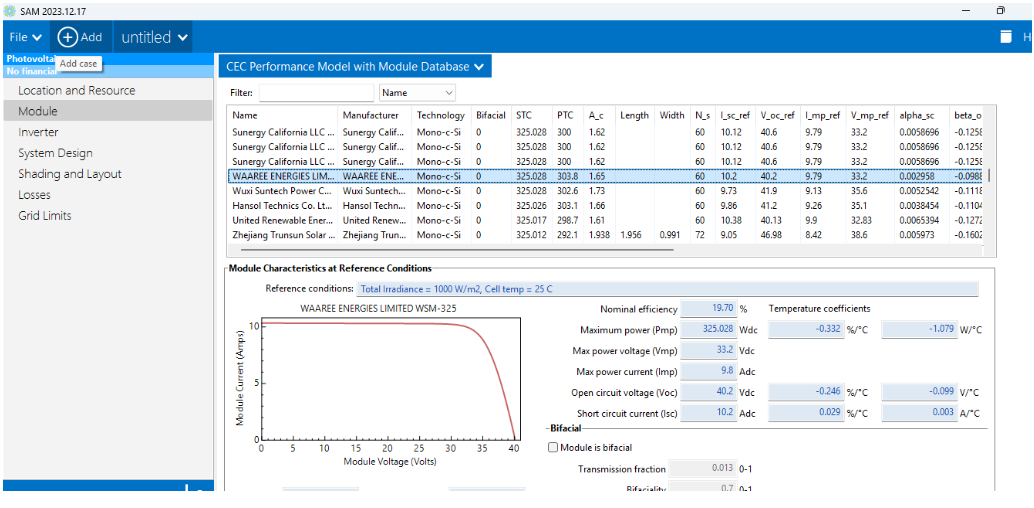}
    \caption{Interfaz de carga de datos SAM.}
    \label{fig:enter-label}
\end{figure}

La imagen muestra la sección de ingreso manual de especificaciones del módulo fotovoltaico en el entorno de SAM, utilizando el modelo de rendimiento CEC. En esta interfaz se detallan parámetros eléctricos clave como el voltaje y la corriente en el punto de máxima potencia (Vmp e Imp), el voltaje en circuito abierto (Voc), la corriente de cortocircuito (Isc), y los coeficientes de temperatura asociados. También se especifican las dimensiones físicas del módulo, el número de celdas en serie y la temperatura nominal de operación. Esta configuración permite personalizar con precisión el comportamiento del panel en el modelo, asegurando su correspondencia con las características reales del sistema instalado.

\begin{figure}[H]
    \centering
    \includegraphics[width=1\linewidth]{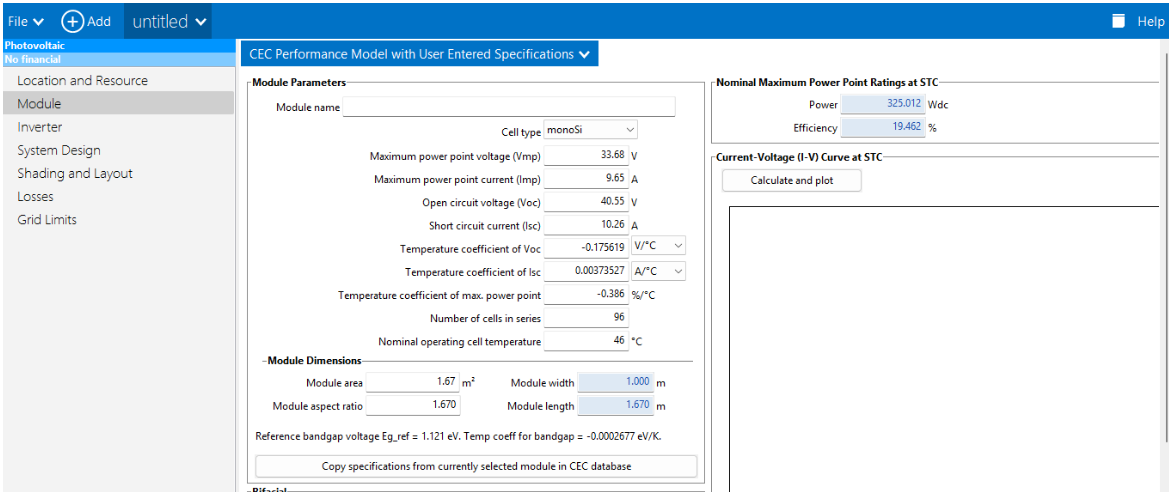}
    \caption{Data entry de las características de los paneles de la instalación en SAM. }
    \label{fig:enter-label}
\end{figure}
Luego, se simularon las producciones mensuales y anuales de energía (kWh) y se generaron gráficos sobre pérdidas, eficiencia del sistema y curvas de rendimiento.
\begin{figure}[H]
    \centering
    \includegraphics[width=1\linewidth]{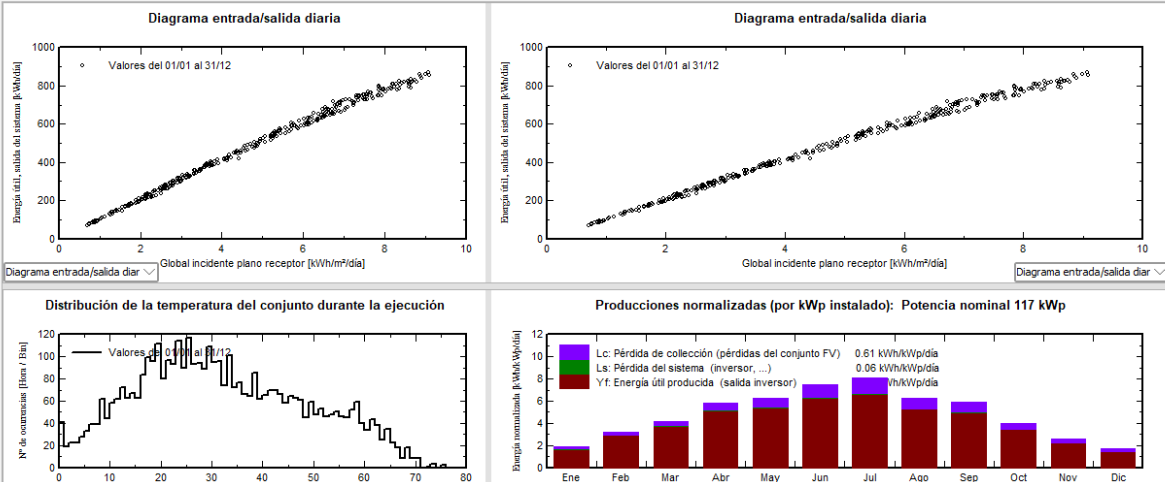}
    \caption{Gráficos de la simulación PV-Syst.}
    \label{fig:enter-label}
\end{figure}
\begin{figure}[H]
    \centering
    \includegraphics[width=1\linewidth]{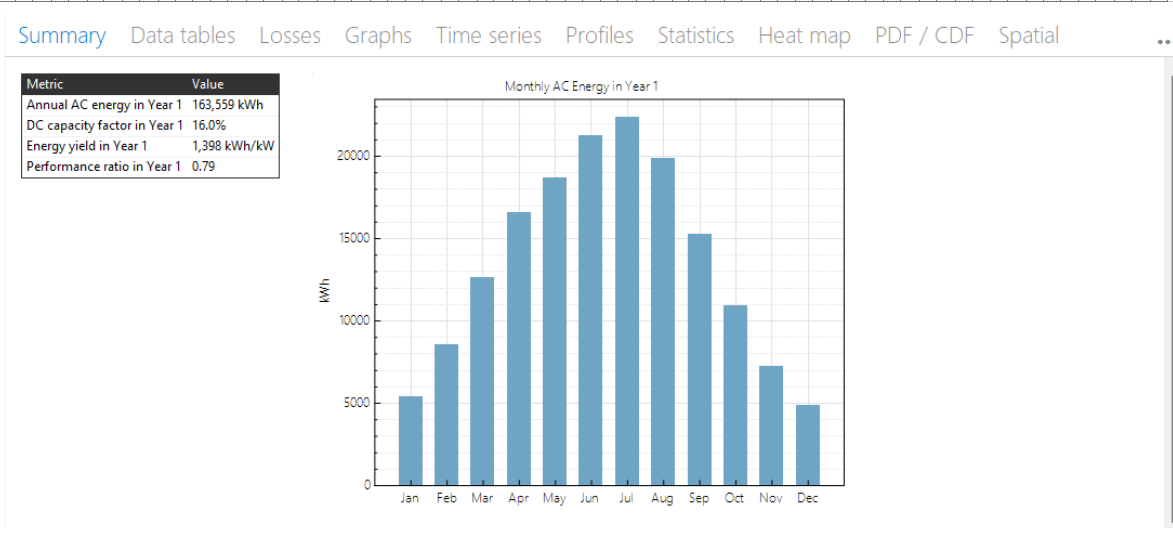}
    \caption{Gráficos de la simulación SAM.}
    \label{fig:enter-label}
\end{figure}
Los datos obtenidos fueron exportados a hojas de cálculo para realizar un análisis cruzado. Finalmente se compararon los resultados simulados con los datos reales de producción registrados en la planta, calculando el porcentaje de desviación mensual y anual. Se analizaron las principales causas de las desviaciones, tales como el modelado de pérdidas, la orientación y la degradación de los paneles.

\subsection{Evaluación del nivel de madurez digital}
En base a la documentación\cite{yuan_digital_2024}, se definieron criterios para clasificar a cada simulador dentro de los niveles DM, DS o DT. Estos son:
\begin{table}[ht]
  \caption{Comparación de criterios evaluados para los niveles digitales DM, DS y DT.}
  \centering
  \renewcommand{\arraystretch}{1.3}
  \begin{tabular}{>{\raggedright\arraybackslash}p{6.5cm} c c c}
    \toprule
    \textbf{Criterio Evaluado} & \textbf{DM} & \textbf{DS} & \textbf{DT} \\
    \midrule
    Importación de datos reales & \checkmark & \checkmark & \checkmark \\
    Actualización dinámica con datos reales & \texttimes & \checkmark & \checkmark \\
    Retroalimentación al sistema físico & \texttimes & \texttimes & \checkmark \\
    Soporte para integración IoT & \texttimes & Parcial & \checkmark \\
    Uso de machine learning & \texttimes & Opcional & \checkmark \\
    Interoperabilidad y scripting & \texttimes & \checkmark & \checkmark \\
    Exportación de resultados & \texttimes & \checkmark & \checkmark \\
    \bottomrule
  \end{tabular}
  \label{tab:comparacion_niveles_digitales}
\end{table}
\section{Resultados}
A continuación, se presentan los gráficos resultantes de las simulaciones, los cuales muestran la comparación entre los datos estimados y los reales mediante el análisis de la desviación cuadrática media (RMSE) y la desviación absoluta media (MAE) para cada mes del año.

\begin{table}[ht]
  \caption{Comparación mensual de energía real y simulada con SAM y PVsyst (en unidades de energía).}
  \centering
  \renewcommand{\arraystretch}{1.2}
  \begin{tabular}{lrrr}
    \toprule
    \textbf{Mes} & \textbf{Real} & \textbf{SAM} & \textbf{PvSyst} \\
    \midrule
    Enero       & 7176,69  & 5382,16  & 6144    \\
    Febrero     & 11576,8  & 8571,03  & 9674    \\
    Marzo       & 14659,21 & 12630,1  & 13705   \\
    Abril       & 16204,36 & 16558,8  & 18359   \\
    Mayo        & 20521,14 & 18672,1  & 20001   \\
    Junio       & 17440,59 & 21214,1  & 22302   \\
    Julio       & 21886,41 & 22341,4  & 24309   \\
    Agosto      & 20740,19 & 19887,6  & 19367   \\
    Septiembre  & 15809,33 & 15286,2  & 17721   \\
    Octubre     & 13567,18 & 10892,6  & 12693   \\
    Noviembre   & 9329,29  & 7261,63  & 7978    \\
    Diciembre   & 6603,83  & 4860,99  & 5270    \\
    \midrule
    \textbf{Total} & \textbf{175515,02} & \textbf{163558,71} & \textbf{177523} \\
    \bottomrule
  \end{tabular}
  \label{tab:energia_mensual}
\end{table}
\begin{figure}
    \centering
    \includegraphics[width=0.7\linewidth]{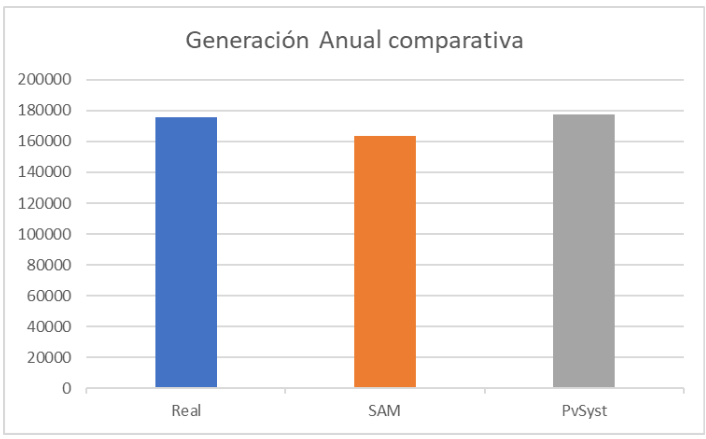}
    \caption{Gráfica comparativa de la generación anual.}
    \label{fig:enter-label}
\end{figure}
\begin{figure}
    \centering
    \includegraphics[width=0.7\linewidth]{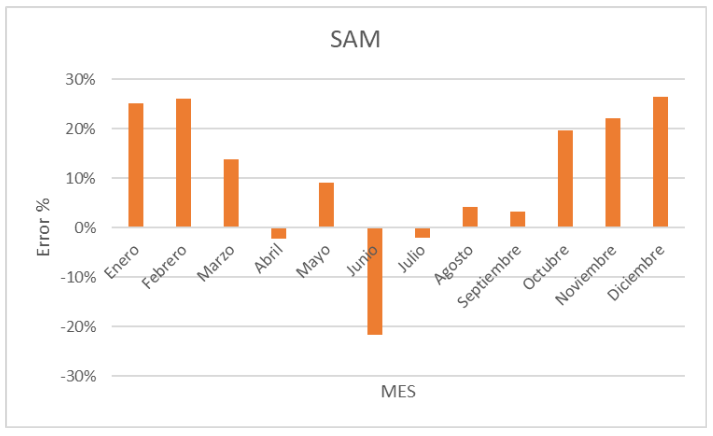}
    \caption{Error porcentual generación SAM Vs Real}
    \label{fig:enter-label}
\end{figure}
\begin{figure}
    \centering
    \includegraphics[width=0.7\linewidth]{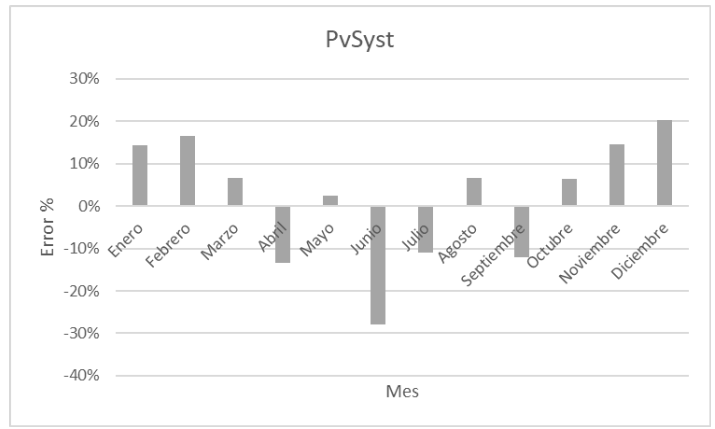}
    \caption{Error porcentual generación Pv-syst Vs Real}
    \label{fig:enter-label}
\end{figure}
\begin{figure}
    \centering
    \includegraphics[width=0.7\linewidth]{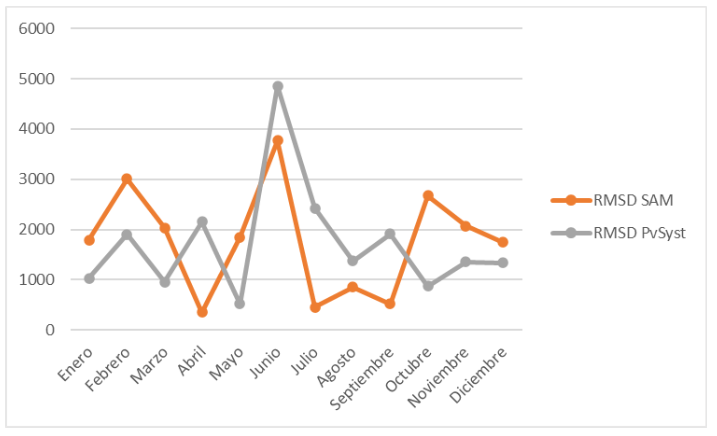}
    \caption{RMSD SAM vs RMSD Pv-Syst}
    \label{fig:enter-label}
\end{figure}
\begin{figure}
    \centering
    \includegraphics[width=0.7\linewidth]{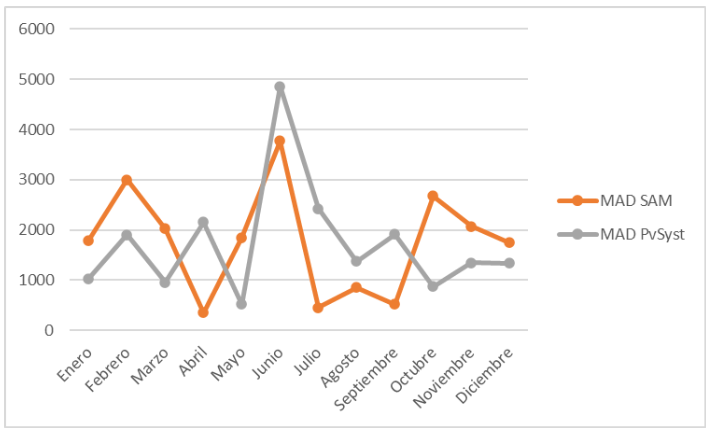}
    \caption{MAD SAM vs MAD PvSyst}
    \label{fig:enter-label}
\end{figure}
\section{Análisis de capacidades digitales}
A partir del marco de clasificación DM/DS/DT y los criterios definidos en la metodología, se evaluó el grado de madurez digital de cada simulador.
\begin{table}[ht]
  \caption{Comparación de características avanzadas entre PVsyst y SAM.}
  \centering
  \renewcommand{\arraystretch}{1.3}
  \begin{tabular}{>{\raggedright\arraybackslash}p{6.5cm} c c}
    \toprule
    \textbf{Característica} & \textbf{PVsyst} & \textbf{SAM} \\
    \midrule
    Importación de datos reales & \checkmark & \checkmark \\
    Actualización dinámica con datos reales & \texttimes & Parcial (vía scripting) \\
    Soporte para integración IoT & \texttimes & Parcial (API externa) \\
    Retroalimentación al sistema físico & \texttimes & \texttimes \\
    Posibilidad de scripting e interoperabilidad & Limitada & Alta \\
    Exportación de resultados para co-simulación & \checkmark & \checkmark \\
    Uso de técnicas de Machine Learning & Bajo & Medio–Alto \\
    \bottomrule
  \end{tabular}
  \label{tab:comparacion_pvsyst_sam}
\end{table}
Cómo se puede observar en la imágen, PVsyst opera, en el mejor de los casos, como un Digital Shadow: es capaz de importar condiciones climáticas históricas y modelar con gran detalle físico el comportamiento del sistema, pero no puede interactuar en tiempo real ni ajustar automáticamente sus parámetros en función del desempeño real del sistema. \cite{wright_digital-twins_nodate} Por otro lado, aunque SAM fue concebido principalmente como una herramienta de análisis técnico-económico, ofrece un mayor nivel de personalización. Gracias a su compatibilidad con scripts en Python y MATLAB, así como la posibilidad de automatizar la entrada y salida de datos, SAM puede evolucionar hacia un entorno más cercano al de un Digital Twin, siempre que se integre con sensores y algoritmos externos.\cite{noauthor_pysam_nodate},\cite{noauthor_how_2023}

\section{Discusión técnica}
La comparación revela que la elección del simulador no solo debe basarse en su precisión técnica, sino en su capacidad de integración en una arquitectura digital más amplia. En un entorno donde la producción energética depende de múltiples variables cambiantes, los simuladores estáticos no son suficientes.

.) PVsyst resulta ideal para la etapa de diseño, donde se busca optimizar la configuración del sistema bajo condiciones promedio o históricas. Su capacidad de análisis de pérdidas y sombreados es superior, pero carece de dinamismo operativo.

.) SAM, en cambio, ofrece una mejor base para la integración en un gemelo digital, gracias a su flexibilidad, sus herramientas de scripting y su compatibilidad con fuentes de datos externas en tiempo real.

Esta observación coincide con los planteamientos de trabajos como el de Guzmán Razo et al. (2023), donde se destaca que la precisión no depende únicamente del modelo físico, sino de la capacidad del sistema de auto ajustarse ante condiciones reales cambiantes.

\section{Implicancias para la implementación de gemelos digitales}
Los resultados permiten afirmar que ninguno de los simuladores analizados constituye por sí solo un gemelo digital completo, pero que SAM tiene mayor madurez estructural para integrarse en una arquitectura de tipo DT\cite{urupira_tafadzwa__rix_arnold_pdf_2017}. Esto implica que:

.)Puede usarse como motor predictivo dentro de un gemelo digital solar.

.)Requiere una capa adicional de adquisición de datos (ej., sensores IoT conectados a plataforma tipo ThingSpeak o InfluxDB).\cite{fernandez_perez_desarrollo_2022}

.)Debe complementarse con módulos de inteligencia artificial.

De este modo, se plantea la necesidad de migrar de una lógica de simulación tradicional (DM) hacia sistemas ciber físicos (DT), donde el modelo, los datos y la inteligencia se retroalimentan continuamente.

\section{Conclusiones}
El desarrollo de este trabajo permitió evaluar de manera crítica el grado de madurez digital de dos de los simuladores fotovoltaicos más utilizados actualmente —PVsyst y SAM— a través de su aplicación a un caso de estudio real. A partir de la taxonomía conceptual de modelos digitales propuesta por Grieves y extendida por Jones et al., se concluye que:

.)PVsyst, aunque altamente preciso en su modelado físico, opera esencialmente como un modelo digital (DM) o, en el mejor de los casos, como una sombra digital (DS), al no poseer capacidades nativas para interactuar dinámicamente con datos en tiempo real ni generar retroalimentación al sistema físico.

.)SAM, en cambio, presenta una arquitectura más flexible, que puede extenderse hacia una configuración DT (Digital Twin), especialmente si se integra con plataformas de monitoreo, APIs y algoritmos de inteligencia artificial. Su capacidad de scripting y acceso a bases de datos climáticas en línea lo posiciona como una opción viable para ser el motor principal de un gemelo digital solar.

El análisis comparativo también evidenció que, si bien ambos simuladores ofrecen estimaciones razonables del rendimiento energético anual, existen limitaciones importantes en su capacidad de adaptación dinámica, lo cual es un requisito esencial en entornos con alta variabilidad climática o demandas energéticas fluctuantes.

\bibliographystyle{unsrt}
\bibliography{Referencias} 
\end{document}